\documentclass[lettersize,journal]{IEEEtran}
\usepackage{amsmath,amsfonts}
\usepackage{algorithmic}
\usepackage{algorithm}
\usepackage{array}
\usepackage[caption=false,font=normalsize,labelfont=sf,textfont=sf]{subfig}
\usepackage{textcomp}
\usepackage{stfloats}
\usepackage{url}
\usepackage{verbatim}
\usepackage{graphicx}
\usepackage{cite}
\usepackage[utf8]{inputenc}
\usepackage{geometry}
\usepackage{url}

\begin{document}

\title{Magnetic Field Induced Ion Trapping Around Dust Particle in Plasma}

\author{N.~Kh.~Bastykova, S.~K.~Kodanova,T.~S.~Ramazanov
\thanks{N. K. Bastykova, S. K. Kodanova, T. S. Ramazanov are with the Institute of Experimental and Theoretical Physics,
Al-Farabi Kazakh National University, Almaty 050040, Kazakhstan (e-mail:
kodanova@physics.kz.)}
\thanks{Manuscript received April 19, 2021; revised August 16, 2021.}}

\markboth{Journal of \LaTeX\ Class Files,~Vol.~14, No.~8, August~2021}%
{Shell \MakeLowercase{\textit{et al.}}: A Sample Article Using IEEEtran.cls for IEEE Journals}

\maketitle

\begin{abstract}
This paper investigates the binary collision between an ion and a charged dust particle in the plasma in the presence of a uniform magnetic field. The trajectories of ions are calculated using the modified Velocity Verlet algorithm designed for a reliable simulation of the charged particle dynamics in arbitrarily strong static homogeneous external magnetic fields. We consider the coupling parameter values characteristic for dusty plasma experiments and the external magnetic field with 
$B\leq 1~{\rm T}$. It has been revealed that at some magnetic field values, the ions are trapped around the charged dust particle. The magnetic field strength values in which the ion becomes trapped around the dust particle have been obtained for typical argon and helium gas discharge plasma parameters.  
\end{abstract}

\begin{IEEEkeywords}
Dusty Plasma, Gas Discharge, Magnetic Field.
\end{IEEEkeywords}

\section{Introduction}
\IEEEPARstart{T}{he} interaction of charged microparticles (dust particles) with other plasma particles is of interest due to the wide field of relevant and perspective applications of dusty plasma in engineering and fundamental research \cite{adamovich20222022,pustylnik2021physical,beckers2023physics, Abdirakhmanov_IEEE_2019}. Generally, macroparticles can cause many different phenomena in laboratory plasmas, such as 
dust acoustic waves \cite{Ludwig2018}, void \cite{morfill1996plasma, barkan1995laboratory, shukla2015introduction} and wakefield formation \cite{Sundar_2020, Moldabekov_cpp_2015}. Charged dusty particles also have a substantial effect on radiation propagation and plasma dynamics in different regions of the solar system associated with space debris, or planetary debris disks near stars \cite{fudamoto2024noema,siebenmorgen2023dark}. In addition, charged dust particles can also be a cause of harmful interferences in the plasma processing of semiconductors \cite{selwyn1989situ}. 

Entering the plasma environment, microparticles acquire a charge and actively interact with plasma electrons and ions. When the concentration of the dust particles is high enough,  it is crucial to understand and adequately take into account the effect of dust particles on plasma properties\cite{hino2002tritium,krasheninnikov2005dynamics,kodanova2015dust, khrapak2003scattering,khrapak2004momentum}.
Once dust particles acquire a charge, a variety of forces, such as electron and ion drag forces, arise due to their collisions with charged plasma particles. These forces are also known to affect the dynamics of dust particles in the plasma. For example, the rotation of dust particles in the magnetized plasma of a gas discharge \cite{abdirakhmanov2021}, as well as the formation of ring-shaped dust structures\cite{kodanova2022ring}, vortexes \cite{abdirakhmanov2023dynamics} and formation of turing patterns \cite{menati2023formation}. In this connection, to understand the physics of these formations it is important to study in detail the pair collisions between the plasma particles and dust particles.
The Orbit Motion Limited (OML) approximation for a spherical dust particle is often used to compute a flux of plasma on a dust particle surface. The OML is applicable in the non-emissive, collisionless and non-magnetized plasmas \cite{goree1994charging}. In addition, collisions of charged particles with neutral particles play an important role in the dust particle charging process and have been successfully theoretically described \cite{khrapak2005particle}. In contrast, the description and understanding of the effect of a strong magnetic field on the plasma-dust particle interaction remain challenging. 
In a strong magnetic field, the charge of dust particles in complex plasmas is successfully computed using the particle-in-cell (PIC) and Monte Carlo methods \cite{Bastykova,kodanova2017effect}. This involves a self-consistent simulation of macroscopic fluxes of electrons and ions on the surface of dust particles. Being accurate and providing a holistic picture, this, however, somewhat masks the peculiarities of the microscopic physics of plasma-dust particle interaction.


In this work, we study a binary collision of an ion with a charged dust particle in a strong external field. This helps to better understand the plasma-dust particle interaction at a microscopic level.
We calculate the dynamics of an ion around a charged dust particle using the Yukawa potential of a point charge. Although simple, this model allows one to gain rather general conclusions about the effect of the external magnetic field on the ion dynamics around the charged dust particle.  As one of the interesting results, we report the trapping of ions in the vicinity of the charged dust particle due to the effect of the magnetic field.
\section{Results.}
We consider the collision of a positively charged ion with a negatively charged dust particle at conditions typical for dusty plasmas generated in gas discharges. The presence of a strong external magnetic field, such as realized in the MDPX (the magnetized dusty plasma experiment device) \cite{Thomas_2020}, results in a strong modification of the orbital motion of the ions around the charged dust particle. In this case, the ion is not only attracted by the electrostatic force to the charged dust particle but also attends a Larmor oscillation. The latter is defined by the velocity of the ion orbiting around the dust particles and the strength of the external magnetic field $B$. We characterize the motion of the ion using the coupling parameter $\beta$ and magnetization parameter $\theta$ defined, respectively, as 

\begin{equation}\label{eq:beta}
 \beta =\frac{e^{2}Z_{i}Z_{d}}{\mu v_{\infty}^2\lambda} ~\mathrm{and}~\theta  = \frac{\Omega_L}{\left(v_{\infty}/ \lambda\right)}=\frac{\lambda}{r_L},\\
\end{equation}
where $v_{\infty}$ is the initial ion velocity, $Z_{d}$ [$Z_{i}$] is the dust particle charge [ion charge],  $\lambda$ is the screening length, $\Omega_L$ ($r_L$) is the Larmor frequency (radius) of the ion moving with the velocity $v_{\infty}$, and $\mu$ is the ion mass. 
  \begin{figure*}[ht]
\centering 
\includegraphics[width=0.7\textwidth]{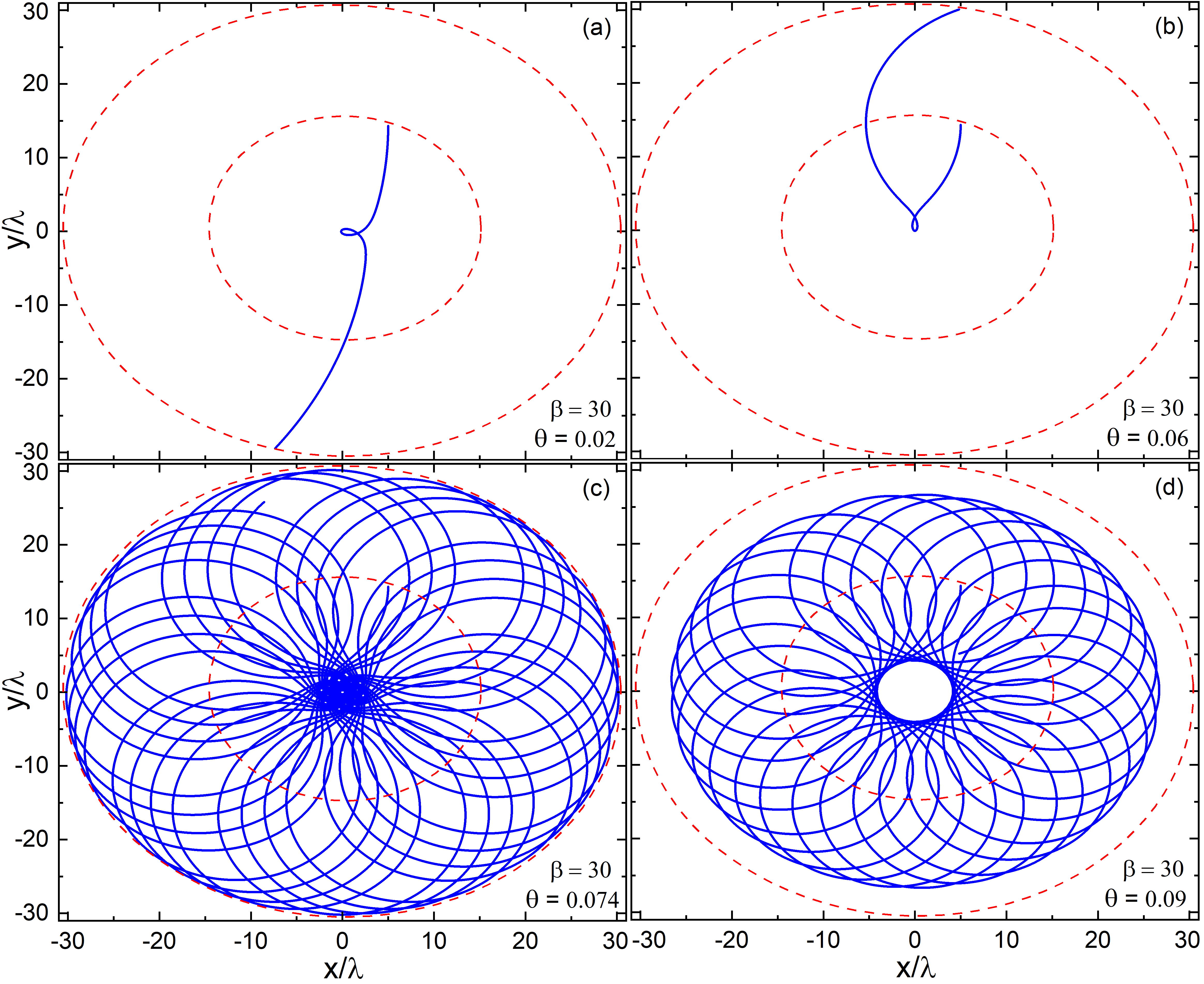}
\centering         \caption{Ion trajectories in the field of the charged dust particle at $\beta = 30$, $\rho/\lambda = 5.0 $, and for a) $\theta = 0.02$, b) $\theta = 0.06$, c) $\theta = 0.074$ and d) $\theta = 0.09$.}  
   \label{fig:1}
\end{figure*}
\begin{figure*}[ht]
\centering 
    \includegraphics[width=0.7\textwidth]{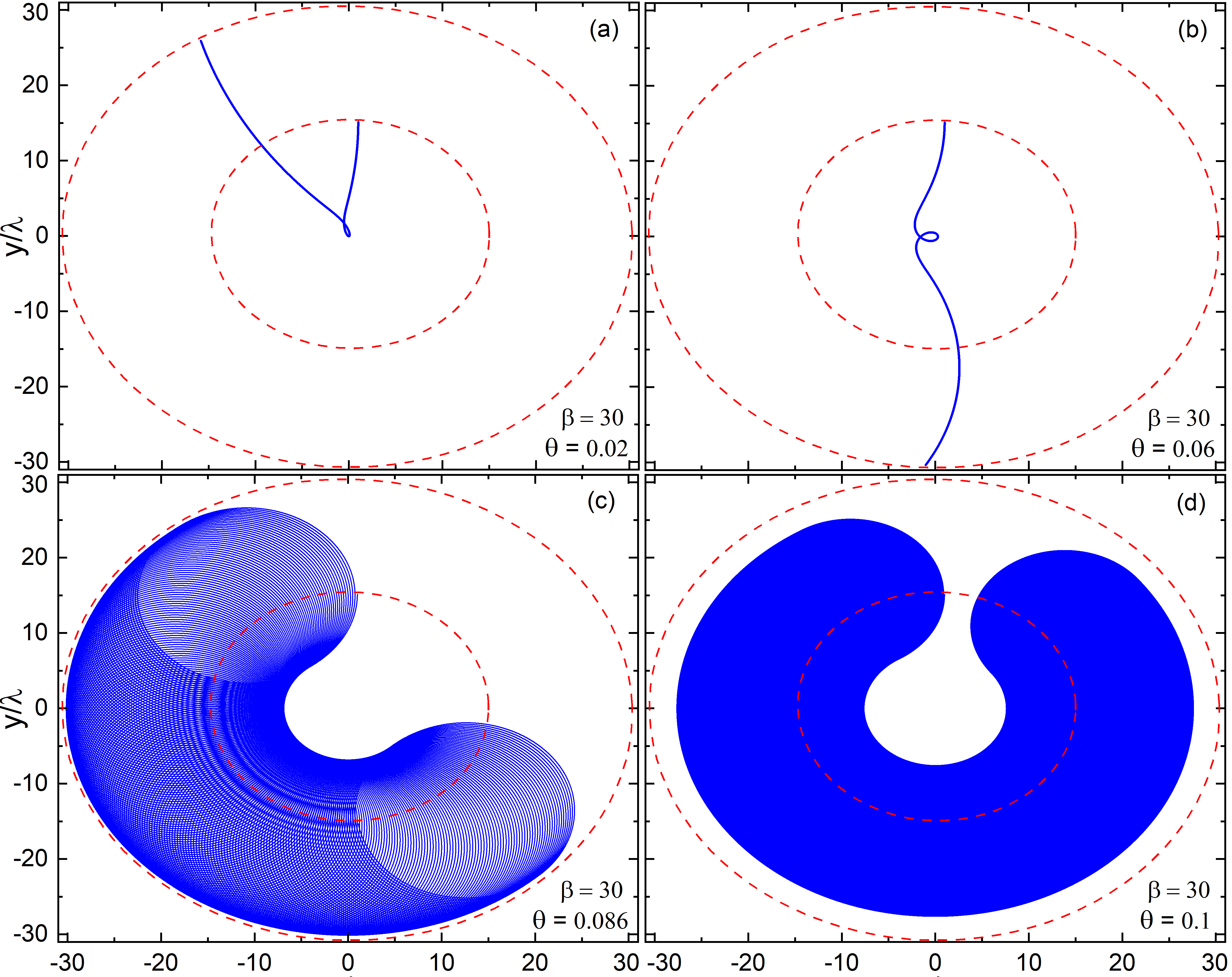}
\centering         \caption{Ion trajectories in the field of the charged dust particle at $\beta = 30$, $\rho/\lambda = 1.0$, and for a) $\theta = 0.02$, b) $\theta = 0.06$, c) $\theta = 0.086$ and d) $\theta = 0.1$.}  
   \label{fig:2}
\end{figure*}

\begin{figure}
\centerline{\includegraphics[width=0.42\textwidth]{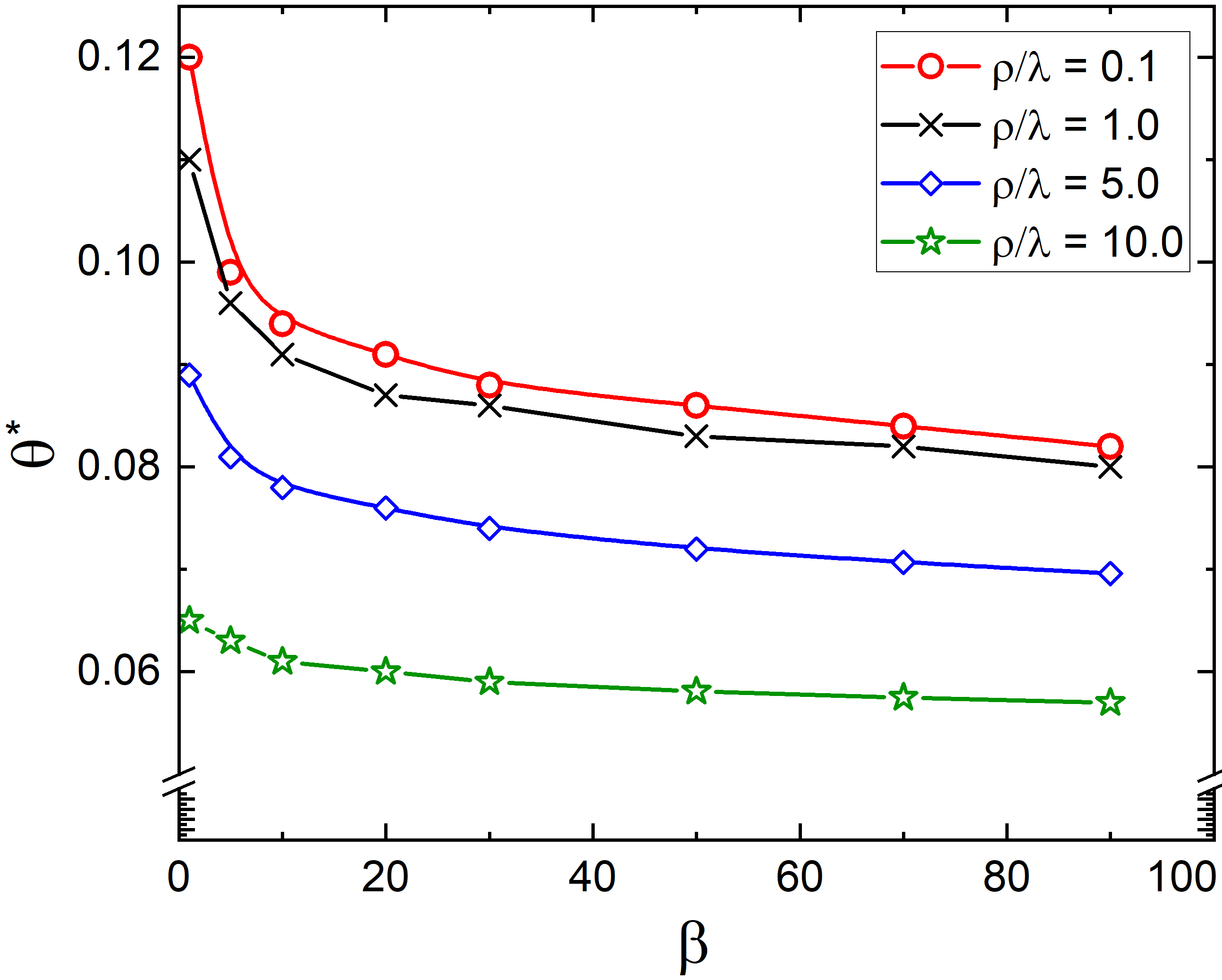}}
        \caption{The dependence of $\theta^*$ on $\beta$ at different impact parameter values $\rho/\lambda$, where $\theta^*$ is the magnetization parameter at which an ion is trapped around a charged dust particle due to the Larmor rotation and orbital motion of the ion around the dust particle as it is illustrated in (c) and (d) subplots of Figs. \ref{fig:1} and \ref{fig:2}.
    \label{fig:3}}
\end{figure}

Using the screening length $\lambda$  as the unit for the distance between the ion and dust particle and $\mu v_{\infty}^2$ as the energy unit, one can write the ion-dust particle interaction potential energy in the following dimensionless form:
\begin{equation}\label{eq:Ueff}
 U(R)=-\frac{\beta}{R}exp(-R). 
\end{equation}

According to potential (\ref{eq:Ueff}), the ion trajectory is calculated using the Verlet algorithm designed for fast and accurate modeling of the charged particle dynamics in arbitrarily strong static homogeneous external magnetic fields  \cite{Spreiter1999ClassicalMD}. In simulations, the magnetic field is taken to be applied along the z-direction.

The calculations of ion motion start at the distance $r_1=r_{cut}$, which is defined by the condition $U(r_{cut})/E=10^{-6}$, where $E=mv_{\infty}^{2}/2$  is the initial (before collision) kinetic energy of the projectile.
The pair collision is considered to be within the sphere with the radius $r_{cut}$.  We follow the ion trajectory until it leaves the region with the radius $r_2= 2r_{cut}$. This allows us to observe the effect of ion capture induced by the magnetic field. This effect is illustrated in Figs. \ref{fig:1} and \ref{fig:2}, where we show the ion trajectories at different values of the magnetization parameter $\theta$ and with the coupling parameter set to $\beta=30$. In Fig \ref{fig:1} we set $\rho/\lambda=5$ and in Fig \ref{fig:2} we use $\rho/\lambda=1$. In Figs. \ref{fig:1} and \ref{fig:2}, the red dashed lines indicate the circles with the radius $r_1=r_{cut}$ and $r_2=2r_{cut}$, and the blue solid line depicts the ion trajectory.

From Figs. \ref{fig:1} and \ref{fig:2}, we see that at $\theta=0.02$ and $\theta=0.06$, the ion leaves the dust particle after the collision. The ion would have been captured on the dust particle surface if the minimal approach distance of the ions was smaller than the dust particle's radius. In Fig. \ref{fig:1} [Fig. \ref{fig:2}],  the increase in the magnetic field strength to $\theta=0.074$ and $\theta=0.09$ [$\theta=0.086$ and $\theta=0.1$] results in the magnetic field-induced rotational motion of the ion around the charged dust particle. In this case, the magnetic field results in the return of the ion into the region with strong ion-dust particle attraction. Consequently, the ion repeats its collision with the dust particle. The ion repeats such motion periodically and becomes effectively captured by the combination of the dust particle's electrostatic field and the uniform external magnetic field.

\begin{figure}
\centerline{\includegraphics[width=0.45\textwidth]{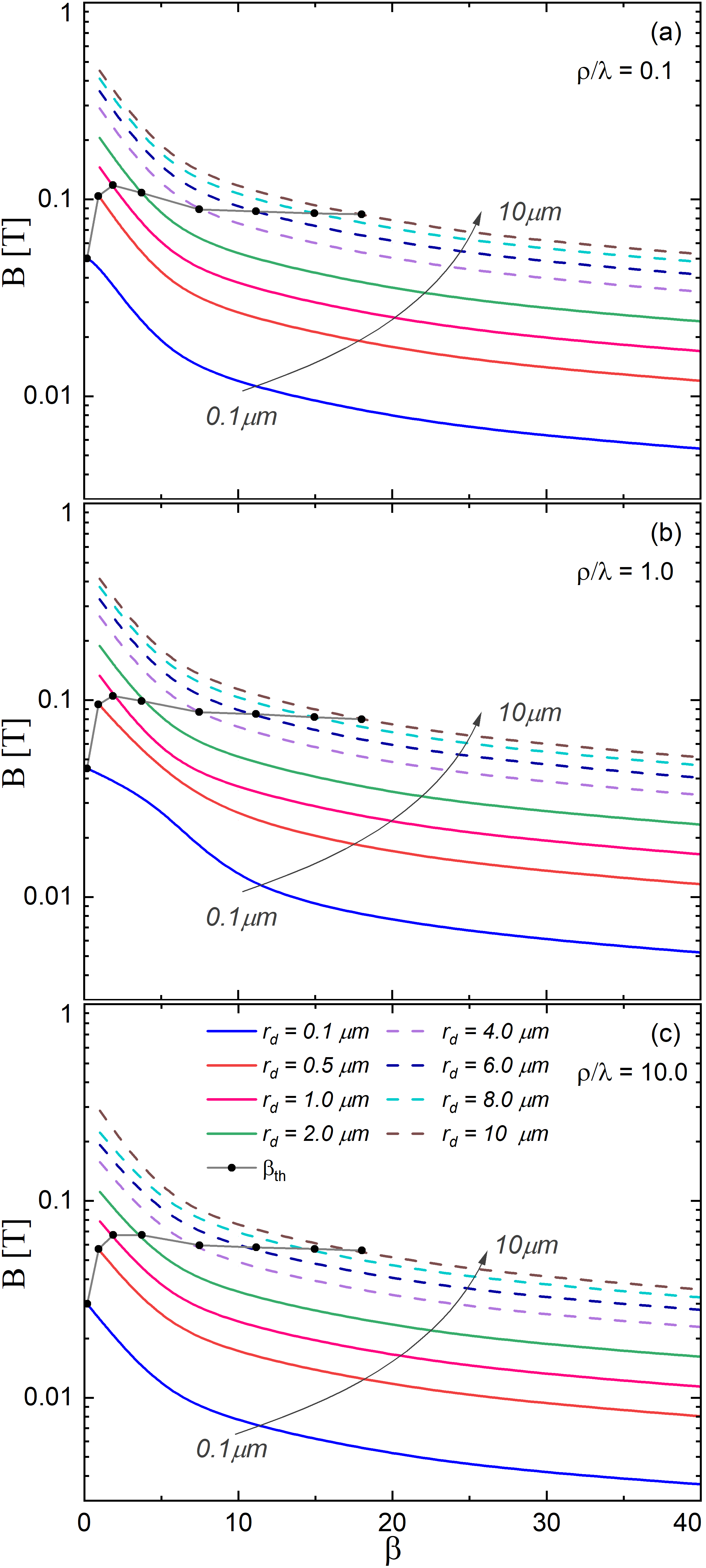}}
        \caption{The dependence of magnetic field strength from $\beta$ at different impact parameter values (a) $\rho/\lambda = 0.1$,(b) $\rho/\lambda = 1.0$, (c) $\rho/\lambda = 10.0$ in helium plasma.
    \label{fig:4}}
\end{figure}
\begin{figure}
\centerline{\includegraphics[width=0.45\textwidth]{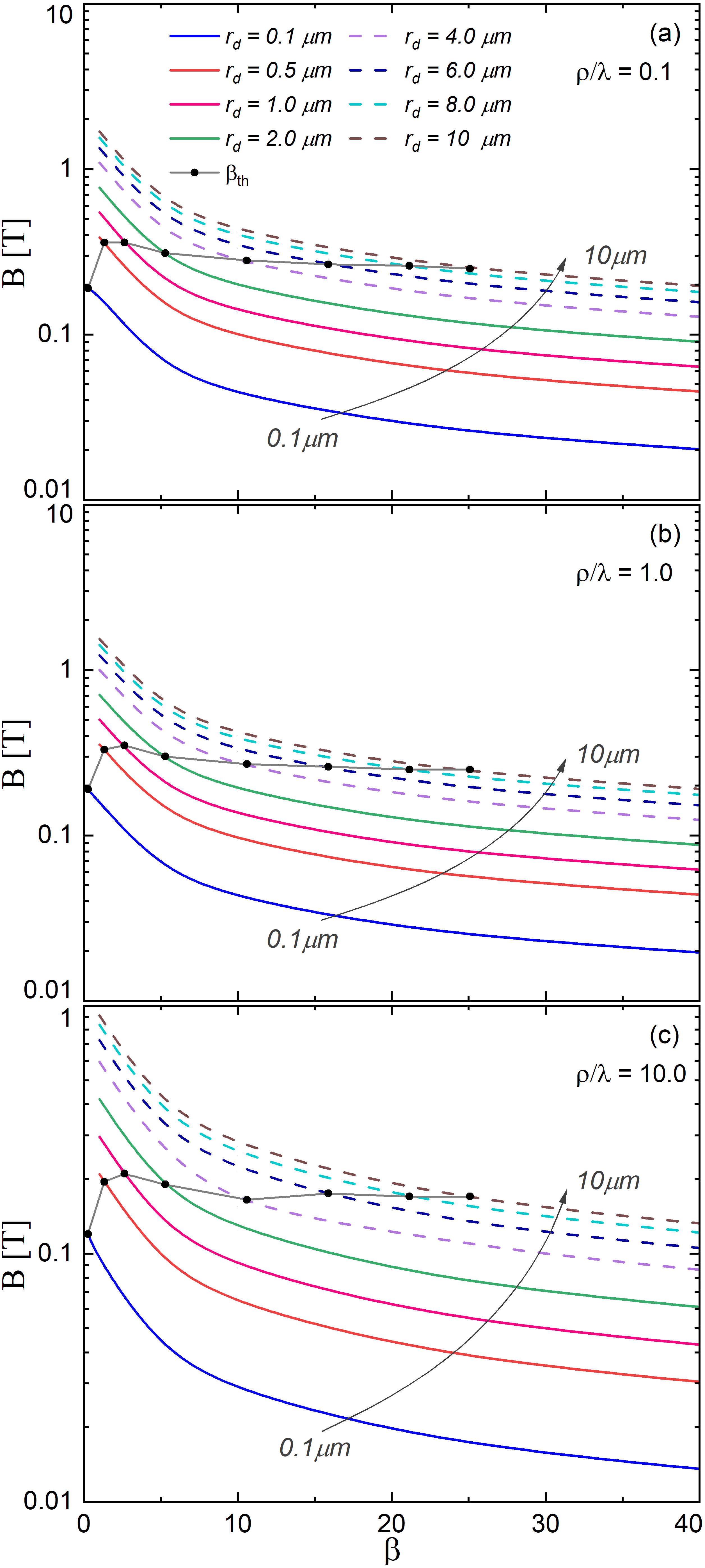}}
        \caption{The dependence of magnetic field strength from $\beta$ at different impact parameter values (a) $\rho/\lambda = 0.1$,(b) $\rho/\lambda = 1.0$, (c) $\rho/\lambda = 10.0$ in argon plasma.
    \label{fig:5}}
\end{figure}

We calculated the trajectory of the ion for the coupling parameter in the range $1\leq \beta \leq 90$, for the impact parameters  $0.1\leq \rho/\lambda\leq 10$, and for the magnetization parameters in the range $10^{-4}\leq \theta \leq 0.12$. 
After analyzing the ion trajectories in the charged dust particle field for different values of the impact parameter  $\rho/\lambda$ and the coupling parameter $\lambda$, the critical value of the magnetization parameter $\theta^{*}$, at which the ion is trapped by the field of the dust particle, has been found. This dependence of $\theta^{*}$ on the coupling parameter $\beta$ at $\rho/\lambda=0.1$, $1.0$, $5.0$, and $10.0 $ is presented in Fig. \ref{fig:3}. As shown in Fig. \ref{fig:3}, the dimensionless parameter $\theta^{*}$ decreases monotonically with the increase in the coupling parameter. At around $\beta<20$, a somewhat faster decrease of $\theta^{*}$ with the increase in $\beta$ is observed compared to the regime with $\beta \geq 20$. The values of $\theta^{*}$ should be considered as an estimate characterizing the peculiarities of the ion capture induced by the external magnetic field. Indeed, the $\theta^{*}$ values depend on the choice of $r_2$, i.e., where we terminate the ion trajectory. However, this does not change the conclusion that the magnetic field, combined with the charged dust particle's electrostatic field, results in the ion's capture in the vicinity of the dust particle in plasmas.


Let us now estimate the characteristic values of the magnetic field strength inducing the capture of an ion around a charged dust particle at helium and argon gas discharge parameters.
We set the density of electrons and ions equal to $n= 10^{15}m^{-3}$, electron temperature $T_e=1 eV$, and ion temperature $T_i=0.027 eV$. These are typical low-pressure gas discharge parameters. In addition, we set $Z_i=1$.
The dust particle radius is varied from $a_d=0.1 \mu m$ to $a_d=10\mu m$.
The magnetic field strength's value as a function of the coupling parameter $\beta$ at different impact parameters (a) $\rho/\lambda = 0.1$,(b) $\rho/\lambda = 1.0$ and (c) $\rho/\lambda = 10.0$ for helium and argon plasma are shown in figures \ref{fig:4}-\ref{fig:5}). The charge of the dust particle for helium and argon plasmas has been calculated using the OML approximation. Accordingly, the value of dust particle charge ranges from $135e$ to $13000e$ for helium plasma and from $191e$ to $18100e$ for argon plasma. The results of the calculations are presented in Fig. \ref{fig:4} for helium plasmas and in Fig. \ref{fig:5} for argon plasmas. 
 As we can see from Figs. \ref{fig:4} and \ref{fig:5}, the increase in the coupling parameter $\beta$ results in smaller magnetic field strength values needed to trap the ion around the dust particle.
 This is in agreement with the results for $\theta^*$ in Fig. \ref{fig:3}. 
 Interestingly, the magnetic field strength values needed to trap an ion increase with the increase in the dust particle radius. This we can understand by recalling that $Z_d\sim a_d$ \cite{melzer2008fundamentals} and $r_L\sim v_{\infty}/B$. Using these relations and Eq. (\ref{eq:beta}), we find that $B\sim \sqrt{a_d}$ for fixed values of $\beta$, $\theta^*$, at given plasma parameters.
 To further clarify the dependence on $\beta$, in Figs. \ref{fig:4} and \ref{fig:5}, we indicate the $\beta(v_{\rm th})=\beta_{\rm th}$ values computed using $v_{\infty}=v_{\rm th}=\sqrt{k_BT_i/\mu}$ in Eq. (\ref{eq:beta}) (depicted using black circles and balk solid lines). From the $B$ values corresponding to $\beta=\beta_{\rm th}$ points, we observe that an ion can be trapped around a charged dust particle with the radius $a_d\geq 0.5$ at $B$ values  order of $0.1~{\rm T}$.
 
\section{Conclusions}
The binary collision between an ion and a charged dust particle in the presence of an external uniform magnetic field has been considered. We analyzed the ion trajectories around the charged dust particle at a wide range of coupling, impact, and magnetization parameters. We report that the external magnetic field can result in the trapping of an ion around a charged dust particle due to the combination of the Larmor rotation and orbital motion of the ion around the dust particle. We evaluated the magnetic field induction values at which such ion trapping can occur in gas discharge plasmas of helium and argon. This insight contributes to a further understanding of the interaction of plasmas with charged dust particles in the presence of an external magnetic field.

\section*{Acknowledgments}
This research is funded by the Science Committee of the Ministry of Education and Science of the Republic of Kazakhstan (Grant BR18574080).


\bibliography{ref}

\end{document}